\documentclass[11pt,a4paper]{article}
\usepackage{jheppub}
\usepackage{subfigure}
\usepackage{tikz}
\usepackage{multirow}
\usepackage{pdflscape}

\usepackage{color}
\definecolor{math1}{rgb}{0.368417,0.506779,0.709798}
\definecolor{math2}{rgb}{0.880722,0.611041,0.142051}
\definecolor{math3}{rgb}{0.560181,0.691569,0.194885}
\definecolor{math4}{rgb}{0.922526,0.385626,0.209179}
\definecolor{math5}{rgb}{0.528488,0.470624,0.701351}

\newcommand{\Tstrut}{\rule{0pt}{2.6ex}}
\newcommand{\Bstrut}{\rule[-0.9ex]{0pt}{0pt}}
\newcommand{\Eqn}[1]{(\ref{#1})}

\newcommand{\D}{\mathrm{d}}

\renewcommand{\vec}{\boldsymbol}

\newcommand{\legerrorpoint}[3]{
\draw[color=#1,fill] (#2) circle (0.05cm) node [right,xshift=5] {\color{black}#3};
\draw[color=#1]      (#2) --+ (0, 0.2) --+ (0.05, 0.2) --+ (-0.05, 0.2);
\draw[color=#1]      (#2) --+ (0,-0.2) --+ (0.05,-0.2) --+ (-0.05,-0.2);
}

\newcommand{\MeV}{\mbox{MeV}}
\newcommand{\GeV}{\mbox{GeV}}

\newcommand{\Em}{E\hspace*{-6pt}/}

\begin{document}
\thispagestyle{empty}
\begin{flushright}
PSI-PR-17-07\\
ZU-TH 11/17\\
\today\\
\end{flushright}
\vspace{3em}
\begin{center}
{\Large\bf Fully differential NLO predictions\\[3pt]
  for the radiative decay of muons and taus}
\\
\vspace{3em}
{\sc G. M. Pruna$^a$, A.~Signer$^{a,b}$, Y. Ulrich$^{a,b}$
}\\[2em]
{\sl ${}^a$ Paul Scherrer Institut,\\
CH-5232 Villigen PSI, Switzerland \\
\vspace{0.3cm}
${}^b$ Physik-Institut, Universit\"at Z\"urich, \\
Winterthurerstrasse 190,
CH-8057 Z\"urich, Switzerland}
\setcounter{footnote}{0}
\end{center}
\vspace{2ex}
\begin{center}
\begin{minipage}[]{0.9\textwidth} {} 
{\sc Abstract:} We present a general purpose Monte Carlo program for
the calculation of the radiative muon decay $\mu\to e\,\nu
\bar{\nu}\gamma$ and the radiative decays $\tau\to e\, \nu
\bar{\nu}\gamma$ and $\tau\to\mu\,\nu \bar{\nu}\gamma$ at
next-to-leading order in the Fermi theory. The full dependence on the
lepton masses and polarization of the initial-sate lepton are
kept. We study the branching ratios for these processes and show that
fully-differential next-to-leading order corrections are important for
addressing a tension between \textsc{BaBar}'s recent measurement of the
branching ratio $\mathcal{B}(\tau\to e\, \nu\bar{\nu}\gamma)$ and the
Standard Model prediction.  In addition, we study various
distributions of the process $\mu\to e\,\nu \bar{\nu}\gamma$ and obtain
precise predictions for the irreducible background to $\mu\to e
\gamma$ searches, tailored to the geometry of the MEG detector.
\end{minipage}
\end{center}

\setcounter{page}{1}

\newpage

\section{Introduction}

The study of muon decays has a very long and rich history and will
continue to play a prominent role in particle
physics~\cite{Kuno:1999jp}. It allows for precise measurements of
Standard Model (SM) input parameters and is a powerful tool in
searches for physics beyond the SM. In this work we are concerned with
the radiative decay of the muon, $\mu\to e\,\nu \bar{\nu}\gamma$. The
branching ratio (BR) for this process depends on the energy cut on the
photon $\omega_0$ that is required to make the quantity well
defined. For the standard choice of $\omega_0 = 10~\MeV$ the BR is
roughly 1\%. Given the vast number of muons that can be produced, it
should be possible to study radiative muon decays with very good
precision.

Despite this, the current experimental knowledge on this process is
rather limited. The standard value that is usually quoted for the
BR is from the sixties and has an error of about
30\%~\cite{Crittenden:1959hm}. There are preliminary results of the
\textsc{Pibeta} collaboration~\cite{Pocanic:2014mya} with much better
precision and MEG has performed a measurement~\cite{Adam:2013gfn} for
rather stringent cuts on the electron and photon energies. Apart from
measuring the BR, the SM can also be tested by measuring Michel
parameters of a general formula for muon
decays~\cite{Eichenberger:1984gi, Pocanic:2014mya,
  Arbuzov:2016ywn}. Furthermore, if the neutrinos have very little
energy, this process is an irreducible background to searches for the
lepton-flavour violating decay $\mu\to e \gamma$.

Related decays of the tau, $\tau\to e\, \nu \bar{\nu}\gamma$ and
$\tau\to \mu\, \nu \bar{\nu}\gamma$ have also been studied. The
measurements of \textsc{Cleo}~\cite{Bergfeld:1999yh} for
$\omega_0=10~\MeV$ (in the rest frame of the tau) have recently been
improved upon by the \textsc{BaBar} collaboration~\cite{Lees:2015gea}
and the BR are now known to about 3\%. Furthermore, a preliminary
analysis concerning the measurement of Michel parameters is also
available from \textsc{Belle}~\cite{Abdesselam:2016fmg}.  Apart from
the usual tests of the SM these decays also offer the possibility to
gain information on the anomalous magnetic moment of the
tau~\cite{Eidelman:2016aih}.

The radiative decay of leptons is most easily computed in the
effective Fermi theory with a four-point interaction.  Corrections
beyond the Fermi theory due to the $W$-boson
propagator~\cite{Ferroglia:2013dga, Fael:2013pja} turn out to be much
smaller than the NLO corrections. The tree-level calculation within
the Fermi theory has been considered by several authors a long time
ago~\cite{PhysRev.101.866, Fronsdal:1959zzb, Eckstein1959297,
  Kinoshita:1959uwa, Fischer:1994pn}. Due to the photon bremsstrahlung
the helicity of the final-state lepton does not have to be
left-handed~\cite{Falk:1993tf,Sehgal:2003mu,Schulz:2004xd,Gabrielli:2005ek}.
After some partial results~\cite{Fischer:1994pn, Arbuzov:2004wr} a
full next-to-leading (NLO) calculation for the BR was presented
in~\cite{Fael:2015gua,Fael:2016hnz}. As for a related calculation of
the rare decays of leptons~\cite{Fael:2016yle}, the results presented
in~\cite{Fael:2015gua,Fael:2016hnz} allow to obtain the differential
decay width at NLO with cuts on the photon and electron energy and
angles between them. In this article, we generalize these results and
present a fully differential Monte Carlo program. This enables the
implementation of arbitrary cuts, allowing to mirror the experimental
situation more closely.

A fully differential calculation has a large impact for radiative
decays of both, the muon and the tau. Starting with the latter, it
has been noted~\cite{Fael:2015gua} that the \textsc{BaBar} result for the BR of
$\tau\to e\,\nu\bar{\nu}\gamma$~\cite{Lees:2015gea} has a tension of
about 3.5 standard deviations with the NLO result. This measurement
was done using rather stringent cuts on the final state and then
converted to the standard cut of $E_\gamma \ge \omega_0 = 10~\MeV$. As
we will show, the NLO corrections are important for this conversion
and potentially resolve this tension. Regarding the muon, taking into
account the geometry of the MEG detector allows to obtain a tailored
background study for $\mu\to e \gamma$ searches.

After briefly discussing some technical aspects of our calculation in
Section~\ref{sec:method} and giving some details about the Monte Carlo
code in Section~\ref{sec:mc} we will study the BR for the rare decays
of the muon and tau in Section~\ref{sec:ratios}.  Whenever possible we
compare with previous results in the
literature~\cite{Fael:2015gua,Fael:2016hnz} and find full
agreement. Making use of the flexibility of our computation, we also
study the impact of the cuts used in the measurement of the BR of the
tau. In Section~\ref{sec:dist} we present several distributions
related to the radiative decay of the muon, adapted to the studies of
MEG and Mu3e. Finally, we present our conclusions in
Section~\ref{sec:conclusion}.

\section{Methodology}\label{sec:method}
This calculation of the radiative decay $\mu\to e\,\nu \bar{\nu}\gamma$
follows very closely the related calculation of the rare decay $\mu\to
e\, \nu \bar{\nu}\,e e$ presented in~\cite{Pruna:2016spf}. The starting
point is the QED Lagrangian with the Fierz rearranged effective
Fermi interaction
\begin{align}
\mathcal{L} &
= \mathcal{L}_{\text{QED}} + \frac{4\, G_F}{\sqrt2} 
\left( \bar\psi_e \gamma^\mu P_L \psi_\mu \right)  
\left( \bar\psi_{\nu_\mu} \gamma^\mu P_L \psi_{\nu_e} \right) + \text{h.c.}\, ,
\label{eq:fierzed}
\end{align}
where $P_L = (1-\gamma_5)/2$ is the usual left-handed projector.  The
QED part of the Lagrangian contains electron $\psi_e$ and muon fields
$\psi_\mu$ but no quark fields. Adding the tau field, $\psi_\tau$ and
making obvious modifications in the four-fermion operator the same
Lagrangian can be used for the radiative decays of the tau.
Despite this being an effective theory, it was shown by Berman and
Sirlin~\cite{BERMAN196220} that all QED corrections are finite after
the usual QED renormalization and that $G_F$ does not get
renormalized. As noted previously, we keep $m_e \neq 0$ and perform
the computation for an arbitrary polarization of the initial-state
lepton. 

At leading order the photon in the radiative decay $\mu\to
e\,\nu\bar\nu \gamma$ can be emitted from the muon or the electron so
that two diagrams need to be calculated. At NLO there are eight
one-loop diagrams, four mass-counterterm diagrams and six real
emissions. The virtual matrix elements are generated using
\texttt{FeynArts}~\cite{Hahn:2000kx} and
\texttt{FormCalc}~\cite{Nejad:2013ina} and evaluated with
\texttt{FORM}~\cite{Kuipers:2012rf} and the scalar integrals were
computed with the library~\texttt{LoopTools}~\cite{Hahn:1998yk}. We
have confirmed the matrix elements with a modified version of
GoSam~\cite{cullen:2014yla,Mastrolia:2012bu, Peraro:2014cba}. The
modification for GoSam necessary to compute polarized amplitudes in
\eqref{eq:fierzed} are
detailed in~\cite{Pruna:2016spf}.

Since we keep all lepton masses different from zero, the phase-space
integration over the real matrix element generates soft, but no
collinear singularities. The former are dealt with using FKS
subtraction~\cite{Frixione:1995ms,Frederix:2009yq}. Since FKS treats
soft and collinear singularities separately, its application in this
case is particularly simple. 

We renormalize the lepton masses and the electromagnetic coupling
$\alpha$ in the on-shell scheme.  The wave-function and the coupling
renormalization factors, as well as the explicit result of the virtual
and real corrections are regularization-scheme dependent. We have used
conventional dimensional regularization (CDR) and the four-dimensional
helicity scheme (FDH)~\cite{Bern:1991aq} and verified that the final
physical result is scheme independent, as described
in~\cite{Signer:2008va}.

We have checked that the corrections beyond the Fermi theory are
suppressed by $m_\mu^2 / m_W^2 \approx10^{-6}$ for $\mu\to e\,\nu\bar\nu
\gamma$ . In particular, as for the normal muon decay, loop diagrams
of the full Standard Model do not introduce logarithmically enhanced
contributions $\log m_\mu^2/m_W^2$.  Hence, for a NLO calculation it
is  sufficient to work within the Fermi theory.

\section{Monte Carlo}\label{sec:mc}

The program presented in this paper is a standard parton-level Monte
Carlo code, or more precisely a Monte Carlo integrator, written in
Fortran~90.  Weighted events with tree-level kinematics (for the Born
term and the virtual corrections) or with an additional photon (for
the real corrections) are generated in the rest frame of the incoming
muon or tau. These events can then be used to define arbitrary
observables and implement arbitrary cuts. The numerical phase-space
integration is done using VEGAS~\cite{Lepage:1980jk}.  The code with
instruction how to use it is available from the authors upon request.

In the case of the real corrections, following the FKS method, we also
have to compute the soft counter event. The FKS procedure introduces
an unphysical parameter, usually denoted by
$\xi_\mathrm{cut}$~\cite{Frixione:1995ms}. This parameter determines
the size of the soft region for which the soft subtraction is carried
out. The dependence of the real corrections on $\xi_\mathrm{cut}$ is
cancelled by the corresponding dependence of the integrated
counterterms. We have checked our implementation by verifying the
independence on $\xi_\mathrm{cut}$ of the full final result, virtual
plus real corrections. Running a particular calculation with different
values of $\xi_\mathrm{cut}$ serves as an indirect check on the
infrared safety of the observables.

Since we keep $m_e\neq 0$ there are no collinear singularities.
However, to ensure numerical stability we have to control pseudo
singularities that arise due to the smallness of $m_e$. To do so we
choose a custom tailored phase-space generator over a recursive
generation to align the pseudo singularities with the variables of the
VEGAS integration.  This optimizes the adaption of the grid for the
integration and leads to very stable numerical results.

Running times obviously depend on the required numerical
precision. The very precise results for the BR given in
Table~\ref{tab:branching} require about 10~CPU~hours per process (on a
reasonably modern machine), whereas a distribution with very precise
numbers for all bins, like the one given in
Figure~\ref{fig:mu3e:photon}, takes of the order of
1000~CPU~hours. Obviously, an arbitrary number of distributions can be
computed in one go and very reasonable results can be obtained
reducing the running time by a factor $5-10$. Typically, about 50\% of
the time is spent on the virtual corrections and the (soft-subtracted)
real corrections each, whereas the integrated counterterm and the Born
term need about 5\% and 1\% of the total time, respectively.

The amount of Monte Carlo statistics necessary to obtain small
numerical uncertainties increases for sharply falling tails of
distributions. To improve the predictions in such cases we configured
special runs that focus on this region. This has been done for the
tail in panel~(a) of Figure~\ref{fig:megdist} where we dedicated a
special run for the region $\Em \le 6~\MeV$. Furthermore, for the
two-dimensional distribution presented in Figure~\ref{fig:meg:double}
the various rows have also been computed in separate runs.

\section{Branching ratios}\label{sec:ratios}

In this section we present some results for the branching ratios
\begin{align}
\mathcal{B}(l,\ell) &= \Gamma(l\to \ell\, \bar\nu\nu \gamma) / \Gamma_l
= \tau_l\ \Gamma(l\to \ell\, \bar\nu\nu \gamma)
\end{align}
with $(l,\ell)\in\{(\tau,\mu),\ (\tau,e),\ (\mu,e)\}$ and $\Gamma_l =
1/\tau_l$ is the experimentally measured width of the lepton. To make
the branching ratio well defined and avoid a  soft
divergence, a minimum photon energy $\omega_0$ has to be
imposed. The usual choice is $\omega_0 = 10~\MeV$.

We have used the following standard values for the various
inputs~\cite{Olive:2016xmw}:
\begin{equation}
\begin{aligned}
m_e &= 0.5109989461(31)~\MeV\,, \\ 
m_\mu &= 105.6583745(24)~\MeV\,, & 
\tau_\mu &= 2.1969811(22)\cdot 10^{-6}\ \mbox{s}\,, \\
m_\tau &= 1776.86(12)~\MeV\,,  & 
\tau_\tau &= 2.903(5)\cdot 10^{-13}\ \mbox{s}\,, \\
G_F &= 1.1663787(6)\times 10^{-11}~\MeV^{-2}\,, & 
\alpha &= 1/137.035999139(31) \, .
\end{aligned}
\end{equation}
Apart from the mean life of the tau they are all known to a precision
that is well beyond what we need.

Table~\ref{tab:branching} summarizes our results for the branching
ratios for the three radiative lepton decays at leading order,
$\mathcal{B}^{\mathrm{LO}}$, and next-to-leading order. In the latter
case we distinguish between the exclusive and inclusive case. For
$\mathcal{B}^{\mathrm{excl}}$ ($\mathcal{B}^{\mathrm{incl}}$) we
require exactly (at least) one photon satisfying the cuts (in this case
energy $E_{\gamma} > \omega_0$).

Let us start by focussing on the first three columns in the top half
of Table~\ref{tab:branching}. There we give the branching ratios with
$\omega_0=10~\MeV$ and no other cuts applied.  Our results agree well
with an earlier calculation presented in~\cite{Fael:2015gua}. The
error given is only the numerical error from the Monte Carlo
integration. 

The size of the NLO corrections for $\mathcal{B}^{\mathrm{excl}}$ are
10\%, 2.5\% and 1.7\% for $(\tau,e)$, $(\tau,\mu)$ and $(\mu,e)$,
respectively.  This is in agreement with the parametric expectation
for the NLO corrections, 
$(\alpha/\pi) \ln(m_l/m_\ell) \ln(\omega_0/m_l)\,
\mathcal{B}^{\mathrm{LO}}$, which amounts to roughly 10\% for $(\tau,e)$
and 3\% for $(\tau,\mu)$ and $(\mu,e)$. Accordingly, we expect the
theoretical error due to omitted NNLO corrections to be about 1\% for
$(\tau,e)$ and considerably smaller for $(\tau,\mu)$ and
$(\mu,e)$~\cite{Fael:2015gua}. As we will see, these estimates are
only valid for a sufficiently inclusive branching ratio. In the case
of the tau, there is an additional error of about 0.2\% due to input
value of the tau mean life.

In order to compare with the recent measurement of
MEG~\cite{Adam:2013gfn} we also present in the last column of
Table~\ref{tab:branching} the $\mu\to e\,\nu\bar\nu\gamma$ branching
ratio for the MEG cuts, i.e. $\omega_0 = 40~\MeV$ and $E_e >
45~\MeV$. Once more, our results agree with an earlier
calculation~\cite{Fael:2016hnz}.  These cuts are very restrictive and
reduce the branching ratio by more than five orders of
magnitude. Furthermore, through these cuts the NLO corrections are
increased from 1.7\% to 5.6\%. Of course, energy conservation forbids
the production of two photons with $E_{\gamma} > 40~\MeV$ together
with $E_e > 45~\MeV$. Hence, there is no distinction between
$\mathcal{B}^{\mathrm{excl}}$ and $\mathcal{B}^{\mathrm{incl}}$ in
this case.

For the radiative decay of the muon, the results presented in
Table~\ref{tab:branching} agree with the experimental values
$\mathcal{B}^{\mathrm{exp}}$~\cite{Crittenden:1959hm, Adam:2013gfn},
albeit with a large experimental error. The situation for the tau
decays is less clear. The most precise measurements of these branching
ratios have been made by
\textsc{BaBar}~\cite{Oberhof2015Measurement,Lees:2015gea}. As argued
in~\cite{Fael:2015gua} these measurements are to be compared with
$\mathcal{B}^{\mathrm{excl}}$. For the $(\tau,\mu)$ decay the
agreement is satisfactory but for the $(\tau,e)$ decay there is a
$3.5\,\sigma$ discrepancy~\cite{Fael:2015gua}. In the remainder of
this section we will revisit this issue, making use of our fully
differential NLO computation to match the actual measurement as
closely as possible.

Let us start with the most problematic case, the decay $\tau\to
e\,\nu\bar\nu\gamma$. For the \textsc{BaBar} measurement, tau pairs are
produced through $e^+ e^-$ collisions at $\sqrt{s} = M_{\Upsilon(4S)}
= 10.58~\GeV$. The event is then divided into a signal- and
tag-hemisphere. In order to reduce background events, rather stringent
cuts on the kinematics of the decay products $e$ and $\gamma$ in the
signal hemisphere are applied. In particular, the following
requirements are made:
\begin{align}
\cos\theta^*_{e\gamma} &\ge 0.97, &
0.22~\GeV &\le E^*_{\gamma} \le 2.0~\GeV, &
M_{e\gamma} &\ge 0.14~\GeV\, .
\label{cut:taue}
\end{align}
All the quantities are given in the centre-of-mass frame. These cuts
can be easily implemented in our code. To this end, we generate taus
in their rest frame, boost them to a frame such that they have energy
$\sqrt{s}/2$ and then apply the cuts \Eqn{cut:taue} in this boosted
frame. As we will see, the NLO corrections will have an important
effect when `undoing' the cuts, i.e. when extracting
$\mathcal{B}^{\mathrm{excl}}$ (with only the cut $E_{\gamma}\ge
10~\MeV$ in the tau rest frame).

In order to illustrate this we have devised the following simplified
scheme: Let $N_{\text{obs}}$ be the measured number of events
including all cuts. To obtain the branching ratio this is multiplied
by a factor $\epsilon_{\text{(N)LO}}^{\text{exp}}$
\begin{align}
\mathcal{B}^{\mathrm{(N)LO}}_{\text{exp}} = \epsilon_{\mathrm{(N)LO}}^{\text{exp}} \cdot
N_{\text{obs}} = 
\epsilon_{\text{det}}\cdot
\epsilon_{\mathrm{(N)LO}} \cdot N_{\text{obs}} \,.
\end{align}
$\epsilon_{\text{det}}$ contains detector efficiencies needed to
compute the fiducial branching ratio. On the other hand,
$\epsilon_{\mathrm{(N)LO}}$ is a theoretical correction factor that is
needed to convert the actually measured branching ratio with the
cuts~\Eqn{cut:taue} to the desired branching ratio with
$E_{\gamma}\ge 10~\MeV$.  This factor can be computed easily at LO
and NLO
\begin{align}
\epsilon_{\mathrm{(N)LO}} =
\frac{\Gamma_{\mathrm{(N)LO}}^{\text{total}}}
  {\Gamma_{\mathrm{(N)LO}}^{\text{with cuts}}} \bigg|_{\rm theory} \,,
\end{align}
where $\Gamma_{\mathrm{(N)LO}}^{\text{total}}$ and
$\Gamma_{\mathrm{(N)LO}}^{\text{with cuts}}$ again refer to the cut
$E_{\gamma}\ge 10~\MeV$ and the cuts~\Eqn{cut:taue},
respectively. More precisely, we require that exactly one photon
passes the cuts. To assess the importance of NLO corrections when
extracting $\mathcal{B}_{\text{exp}}$ we write
\begin{align}
\mathcal{B}^{\mathrm{NLO}}_{\text{exp}}
= \epsilon_{\mathrm{NLO}}^{\text{exp}} \cdot N_{\text{obs}}
= \epsilon_{\mathrm{NLO}} \cdot \epsilon_{\text{det}}\cdot N_{\text{obs}}
= \frac{\epsilon_{\mathrm{NLO}}}{\epsilon_{\mathrm{LO}}} \cdot 
  \mathcal{B}^{\mathrm{LO}}_{\text{exp}}
= \epsilon' \cdot \mathcal{B}^{\mathrm{LO}}_{\text{exp}} \,.
\end{align}
Thus, $\epsilon'$ is a purely theoretical factor that describes the
difference of using a LO or NLO computation in the determination of
$\mathcal{B}_{\text{exp}}$. 

The results for the various factors described above are given in the
first row of Table~\ref{tab:branching}. The salient feature is that
NLO effects are very important in the $\tau\to e\,\nu\bar\nu\gamma$ case
and amount to a correction of 7\%. Since the corresponding
\textsc{BaBar} result was obtained using theory at LO the inclusion of
the NLO corrections changes the result from $\mathcal{B}_{\text{exp}}
= 1.847(54)\cdot 10^{-2}$ to $\epsilon' \cdot \mathcal{B}_{\text{exp}}
= 1.704(50)\cdot 10^{-2}$, in much better agreement with the
theoretical NLO result $\mathcal{B}^{\text{excl}} = 1.645(1)\cdot
10^{-2}$. We recall that we expect an additional theoretical error of
the order of 1\% due to the uncomputed higher-order corrections.

Of course, the same procedure can be repeated for the
$\tau\to\mu\,\nu\bar\nu\gamma$ decay. In this case, some of the cuts
applied by \textsc{BaBar} are 
\begin{align}
\cos\theta^*_{\mu\gamma} &\ge 0.99, &
0.10~\GeV &\le E^*_{\gamma} \le 2.5~\GeV, &
M_{\mu\gamma} &\le 0.25~\GeV\, .
\label{cut:taumu}
\end{align}
A computation of the $\epsilon'$ factor reveals that the effects here
are more modest and amount only to a correction of about 1\%. The
resulting value $\epsilon' \cdot \mathcal{B}_{\text{exp}} = 3.65(10)
\cdot 10^{-3}$ agrees well with the NLO result
$\mathcal{B}^{\text{excl}} = 3.571(1)\cdot 10^{-3}$.

For completeness, we also apply a similar procedure to the MEG result
for the branching ratio of $\mu\to e\,\nu\bar\nu\gamma$. The cuts given
in \Eqn{cut:mue} are applied in the rest frame of the muon. As can be
seen from the result in the last column of Table~\ref{tab:branching},
in this case the NLO effects are marginal and it is sufficient to use
a LO description to remove the effect of the angular cuts in
\Eqn{cut:mue}.

\begin{figure}[t]
\centering
\begin{tabular}{c||c|c|c||c}
      & $\tau\to e\,\bar\nu\nu \gamma$ & $\tau\to \mu\,\bar\nu\nu\gamma$ & $\mu \to e \bar\,\nu\nu \gamma$      & $\mu\to e\,\bar\nu\nu\gamma$  \\\hline
$\mathcal{B}_{\mathrm{LO}}$ & $1.834(1)\cdot 10^{-2}$     & $3.662(1)\cdot 10^{-3}$      & $1.308(1)\cdot 10^{-2}$    &$6.203(1)\cdot 10^{-8}$ \Tstrut     \\
$\mathcal{B}^{\text{excl}}$& $1.645(1)\cdot 10^{-2}$     & $3.571(1)\cdot 10^{-3}$      & $1.286(1)\cdot 10^{-2}$   & \multirow{2}{*}{$5.850(1)\cdot 10^{-8}$}    \\
$\mathcal{B}^{\text{incl}}$& $1.727(3)\cdot 10^{-2}$     & $3.604(1)\cdot 10^{-3}$      & $1.289(1)\cdot 10^{-2}$      \\
\hline
$\mathcal{B}_{\text{exp}}$ & $1.847(54)\cdot 10^{-2}$     & $3.69(10)\cdot 10^{-3}$      & $1.4(4)\cdot 10^{-2}$              & $6.03(55)
\cdot 10^{-8}$   
\Tstrut\Bstrut \\\hline\hline
$\epsilon_{\mathrm{ LO}}  $ & $48.55(1)$  & $4.966(1)$ &   & $9.624(2)$    \\
$\epsilon_{\mathrm{NLO}}  $ & $44.80(1)$  & $4.911(1)$ &   & $9.619(2)$    \\
$\epsilon' =   \epsilon_{\mathrm{NLO}}/\epsilon_{\mathrm{LO}}$ 
                         & $0.923(1)$  & $0.989(1)$ &  & $0.9995(3)$    \\
$\epsilon' \cdot
 \mathcal{B}_{\text{exp}}$ & $1.704(50)\cdot 10^{-2}$   & $3.65(10) \cdot 10^{-3}$   &     & $6.03(55) \cdot 10^{-8}$    \\
\end{tabular}
\renewcommand{\figurename}{Table}
\caption{ Branching ratios for the radiative decays of $\tau$ and
  $\mu$ leptons.  Except in the last column the minimum energy is
  $\omega_0 = 10~\MeV$.  In the last column we used the MEG cuts, i.e.
  $\omega_0 = 40~\MeV$ and $E_e > 45~\MeV$. For the theoretical
  results only the numerical error due to the Monte Carlo integration
  is given. The errors on the experimental results are combined
  statistical and systematic errors, as given by the
  collaborations~\cite{Lees:2015gea, Crittenden:1959hm, Adam:2013gfn}.
}
\label{tab:branching}

\end{figure}

Obviously, this is only a simplistic and by far not complete
simulation of the full analysis.  While the cut on $E^*_{\gamma}$ has
the biggest impact, the results for the $\epsilon'$ factor actually
depend quite significantly on all the details of the cuts.  In
particular, in the presence of a second photon it is important to
precisely specify how the cuts are applied.  This can also be seen
from the rather large difference between $\mathcal{B}^{\text{excl}}$
and $\mathcal{B}^{\text{incl}}$ for $\tau\to e\,\nu\bar\nu\gamma$.  We
do not claim that this is the conclusive resolution to the apparent
$3.5~\sigma$ deviation for the measured branching ratio of $\tau\to
e\,\nu\bar\nu\gamma$. However, we do claim that a proper inclusion of
NLO effects is mandatory for such a measurement, in particular if
stringent cuts on the decay products are applied.

Finally, we consider the branching ratio for $\mu\to
e\,\nu\bar\nu\gamma$ with $\omega_0 = 10~\MeV$ and an additional cut on
the angle between the electron and the photon, $\theta_{e\gamma} >
30^\circ$. For this BR, a preliminary result
$\mathcal{B}_{\text{exp}}=4.365(43) \cdot 10^{-3}$ is
available~\cite{Pocanic:2014mya} from the \textsc{Pibeta}
collaboration. We find\footnote{We thank M.~Fael and M.~Passera for
confirming our results in \eqref{res:pibeta}.}
\begin{align}
\label{res:pibeta}
\mathcal{B}_{\mathrm{LO}} &= 4.264(1)\cdot 10^{-3}\, , & &
\mathcal{B}^{\text{excl}} = 4.228(1)\cdot 10^{-3}\, ,& &
\mathcal{B}^{\text{incl}} = 4.233(1)\cdot 10^{-3}\, .
\end{align}
The error in \eqref{res:pibeta} is the numerical error from the Monte
Carlo integration only. Following the arguments given above, a
conservative estimate of the theory error due to higher-order
correction is $\lesssim 0.2\%$.  There seems to be a tension between
theory and experiment. However, the experimental result is only
preliminary and the SM value quoted by the collaboration disagrees
significantly with our result. Hence, further clarification is
required before a conclusion can be drawn.

\section{Distributions}\label{sec:dist}

While branching ratios with only an $\omega_0$ cut are useful standard
quantities, differential distributions with arbitrary cuts offer a
more direct comparison between experiment and theory. Furthermore,
they are important for precise background estimation. As we have seen
in the previous section, not taking into account the cuts
systematically at NLO can have surprisingly large effects. 

In this section we will provide some examples of such distributions
that can be created with our Monte Carlo program. We will restrict
ourselves to the radiative decay of the muons, but obviously similar
results can be obtained for radiative decays of the taus.

A very promising place to study the process $\mu\to e\,\nu\bar\nu\gamma$
is the MEG experiment~\cite{Baldini:2013ke}. Because it was designed
to search for LFV muon decays its detector geometry creates a very
restrictive phase space. We mimic the MEG detector by imposing the
following cuts:
\begin{subequations}
\label{cut:mue}
\begin{align}
E_\gamma > 40\,\MeV\,,\qquad
&E_e     > 45\,\MeV
               \label{eq:cuts:energy}
\,,\\
|\cos\sphericalangle(\vec p_\gamma, \vec z)|
 \equiv|\cos\theta_\gamma| < 0.35\,,\qquad
&      |\phi_\gamma|       > \frac{2\pi}3 
               \label{eq:cuts:geom_photon}
\,,\\
|\cos\sphericalangle(\vec p_e, \vec z)|
 \equiv|\cos\theta_e     | < 0.5\,,\qquad
&      |\phi_e           | < \frac{\pi}3
               \label{eq:cuts:geom_electron}\,.
\end{align}
\end{subequations}
The muon polarization is set to be $85\%$ along the $\vec z$-axis such
that $\vec P_\mu = -0.85\ \vec z$.

In addition, we require one and only one photon, meaning that an event
with additional real radiation with energy larger than the detector
resolution of roughly $2\,\MeV$ will be discarded if it hits the
detector, i.e. also follows~\eqref{eq:cuts:geom_photon}.
\begin{align}
E_{\gamma_2} < \begin{cases}
2\,\MeV & \text{if \eqref{eq:cuts:geom_photon} is satisfied}\\
\infty & \text{otherwise}
\end{cases}
\tag{\theequation d} \,.\label{eq:cuts:soft}
\end{align}
This approximation is justified even though it would reject a pair of
collinear photons because, in contrast to QCD, there is no mechanism
preferring collinear photons. 

\begin{figure}[t]
\centering
\begin{minipage}[b]{0.49\textwidth}
\subfigure[The polar electron distribution]{
\scalebox{0.7}{\begin{tikzpicture}
\node [anchor=south] at (0,0) {\includegraphics[width=1.3\textwidth]{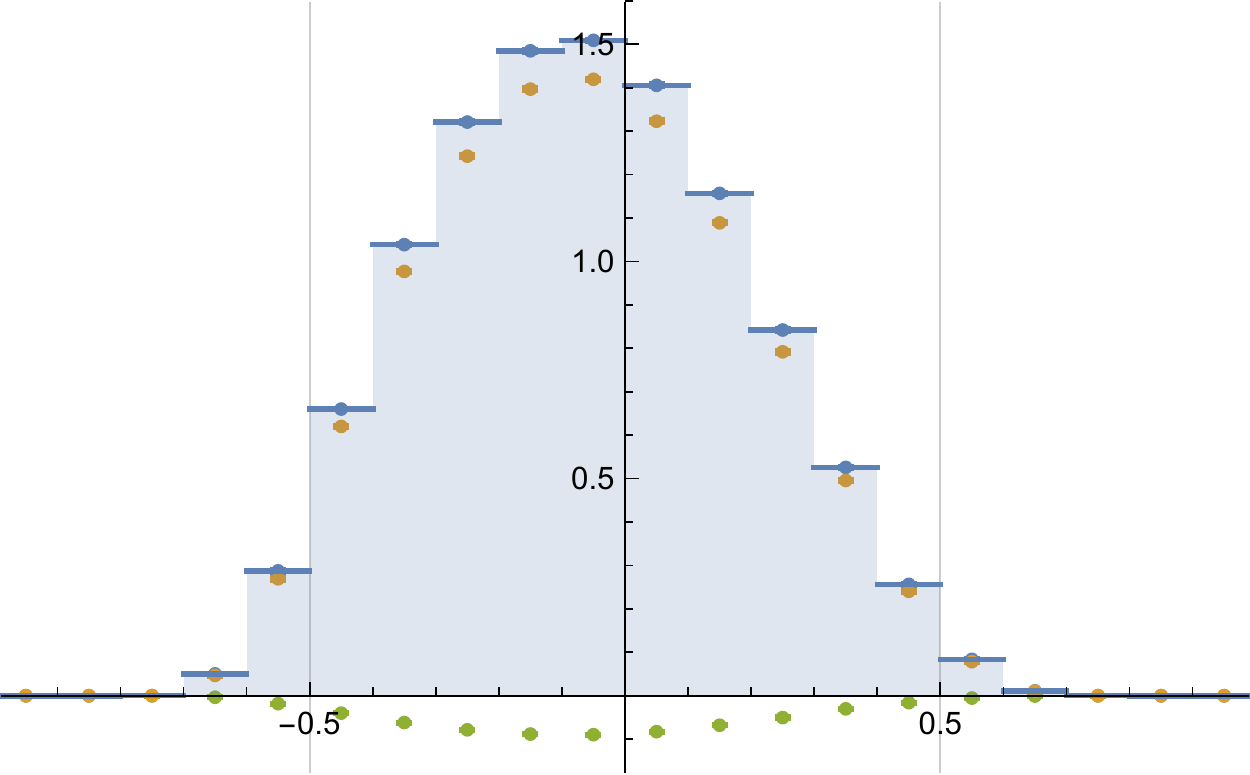}};
\node at (4.5,1) {$\cos\theta_e$};
\node at (0,6.5) {$\frac1{\mathcal{B}_{\text{nb}}}\frac{\D\mathcal{B}}{\D \cos\theta_e}$};
\end{tikzpicture}
}}
  \end{minipage}
  \hfill
  \begin{minipage}[b]{0.49\textwidth}

\subfigure[The azimuthal electron distribution]{
\scalebox{0.7}{
\begin{tikzpicture}
\node [anchor=south] at (0,0) {\includegraphics[width=1.3\textwidth]{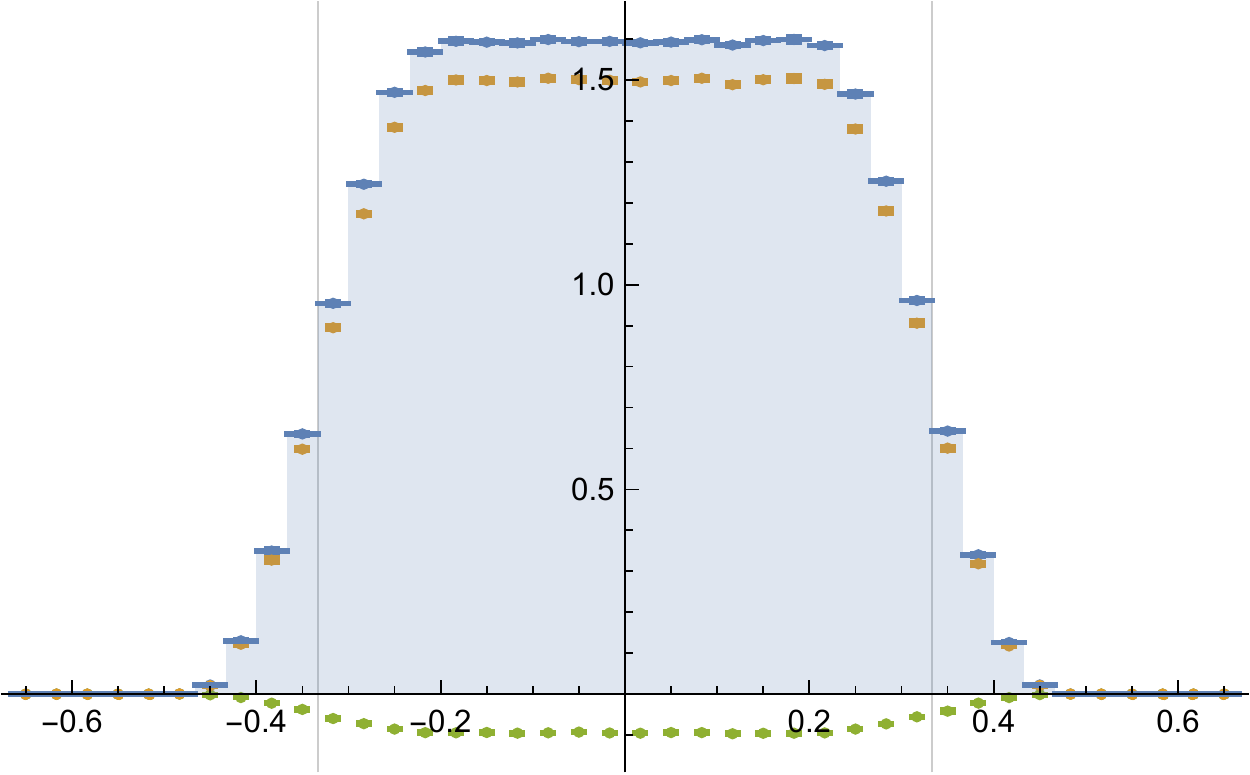}};
\node at (4.5,1) {$\phi_e / \pi$};
\node at (0,6.5) {$\frac1{\mathcal{B}_{\text{nb}}}\frac{\D\mathcal{B}}{\D\phi_e}$};
\end{tikzpicture}
}}
  \end{minipage}

\caption{The angular distributions of the electron at LO (blue), NLO
  (orange) and the correction (green) normalized to NLO result
  $\mathcal{B}$ with the cuts \eqref{eq:cuts:energy},
  \eqref{eq:cuts:geom_photon} and \eqref{eq:cuts:soft}.  The cuts
  \eqref{eq:cuts:geom_electron} are shown in grey lines.}

\label{fig:meg:ang}
\end{figure}
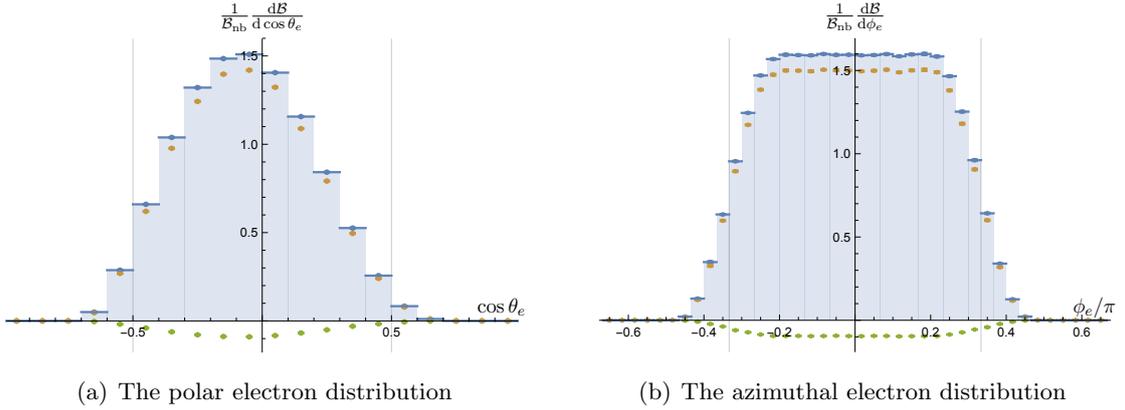

The cuts enforced by the finite size of the electron
detector~\eqref{eq:cuts:geom_electron} would  automatically be
satisfied indirectly through the photon
cuts~\eqref{eq:cuts:geom_photon}, if they were emitted back-to-back.
In order to analyse the loss of events that is due to this
approximation we consider the distributions
\begin{align}
\frac{\D\mathcal{B}}{\D(\cos\theta_e)}\quad
\text{and}\quad\frac{\D\mathcal{B}}{\D\phi_e}\,,
\end{align}
without implementing the cuts~\eqref{eq:cuts:geom_electron}. These
distributions of the branching ratio $\mathcal{B}$ are shown in
Figure~\ref{fig:meg:ang}.  Integrating them at NLO in the interval
specified by~\eqref{eq:cuts:geom_electron} we find
\begin{align*}
\frac1{\mathcal{B}}\int_{-1/2}^{1/2}\D(\cos\theta_e)\  
\frac{\D\mathcal{B}}{\D(\cos\theta_e)}  = 0.959
\quad\text{and}\quad
\frac1{\mathcal{B}}\int_{-\pi/3}^{\pi/3}\D\phi_e\   
   \frac{\D\mathcal{B}}{\D\phi_e}  = 0.929\,.
\end{align*}
Hence the $\theta_e$ and $\phi_e$ cuts result in a loss of about 4\%
and 7\%, respectively. Due to the non-vanishing polarization, the
$\cos\theta_e$ distribution is not symmetric w.r.t. zero. We also see
that the corrections are negative and amount to roughly 5--10\%. As we
will see, this is a generic feature for the distributions, except for
special corners of phase space.

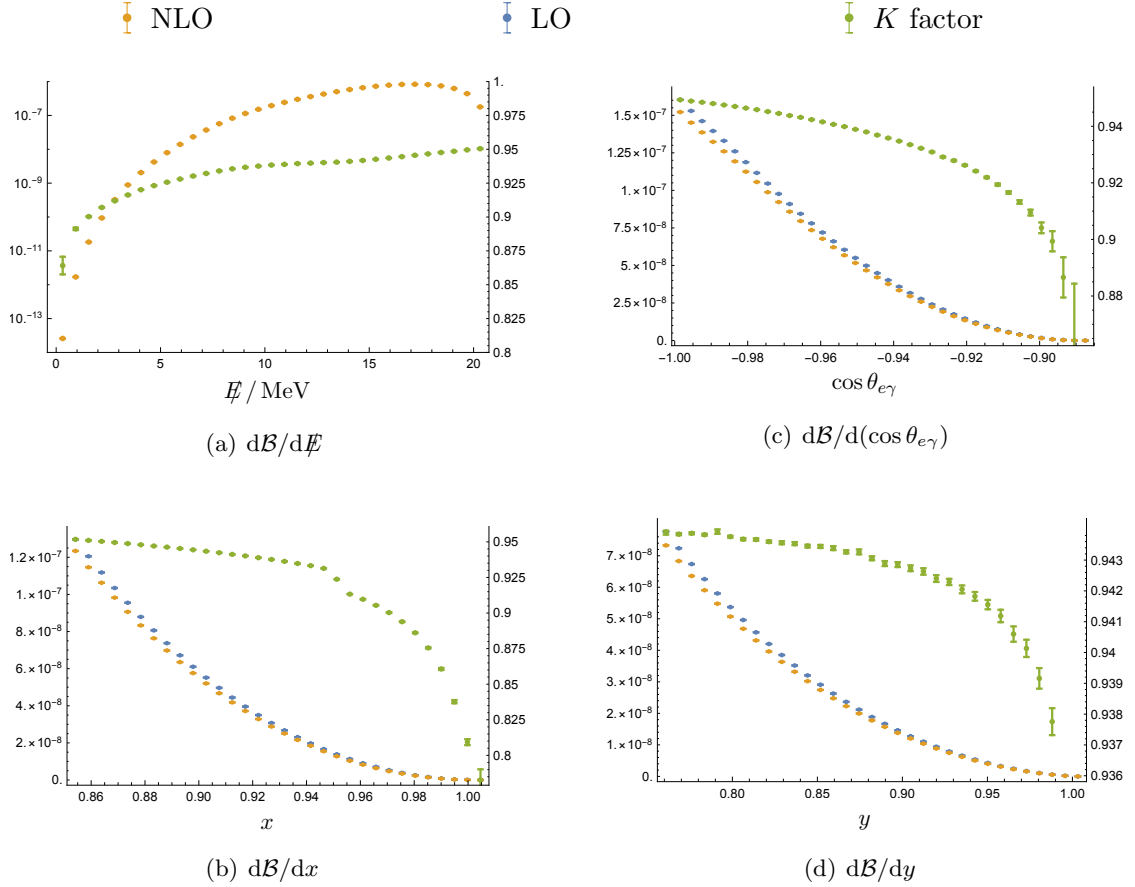
\begin{figure}[t]
\centering

\begin{tikzpicture}
\legerrorpoint{math2}{-7cm,0}{NLO }
\legerrorpoint{math1}{-2cm,0}{LO }
\legerrorpoint{math3}{2.5cm,0}{$K$ factor}
\end{tikzpicture}
\\[10pt]

\begin{minipage}[b]{0.48\textwidth}
\subfigure[$\D\mathcal{B}/\D \Em$]{
\scalebox{0.8}{
\begin{tikzpicture}
\node [anchor=south] at (0,0) {\includegraphics[width=8.5cm]{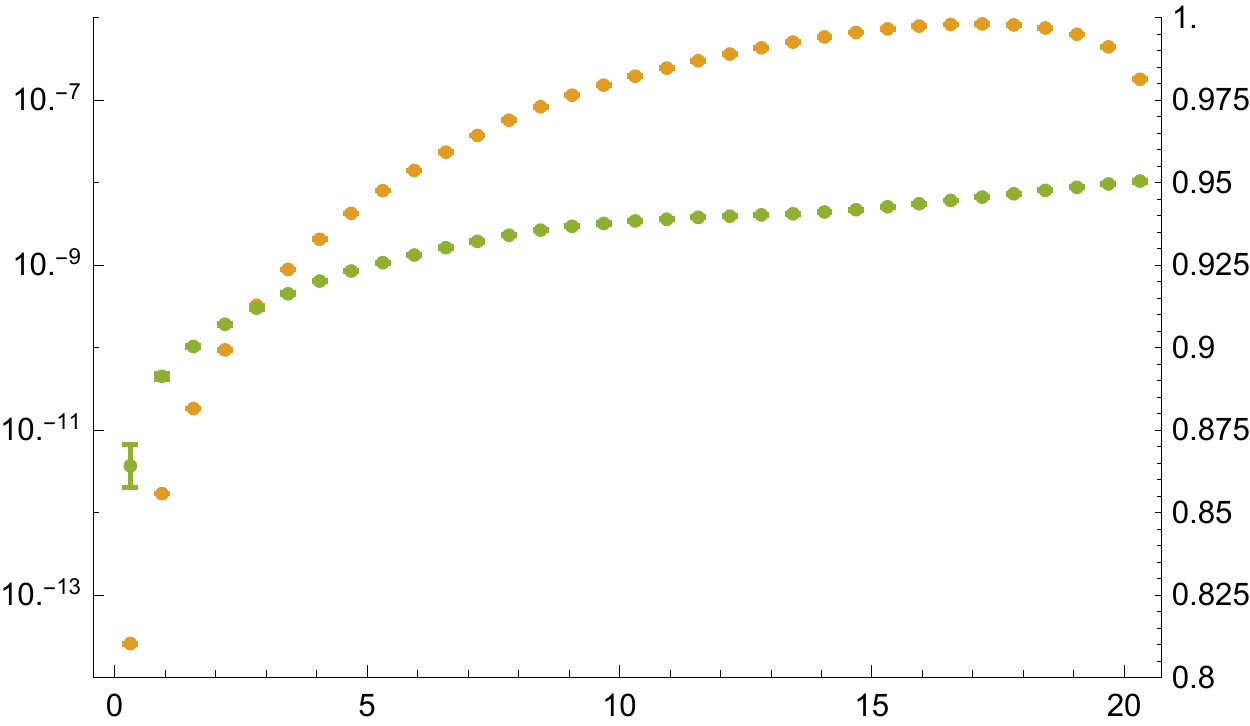}};

\node at (0,-0.2) {$\Em\, /\, \MeV$};

\end{tikzpicture}
}}

\bigskip

\subfigure[$\D\mathcal{B}/\D x$]{
\scalebox{0.8}{
\begin{tikzpicture}
\node [anchor=south] at (0,0) {\includegraphics[width=8.5cm]{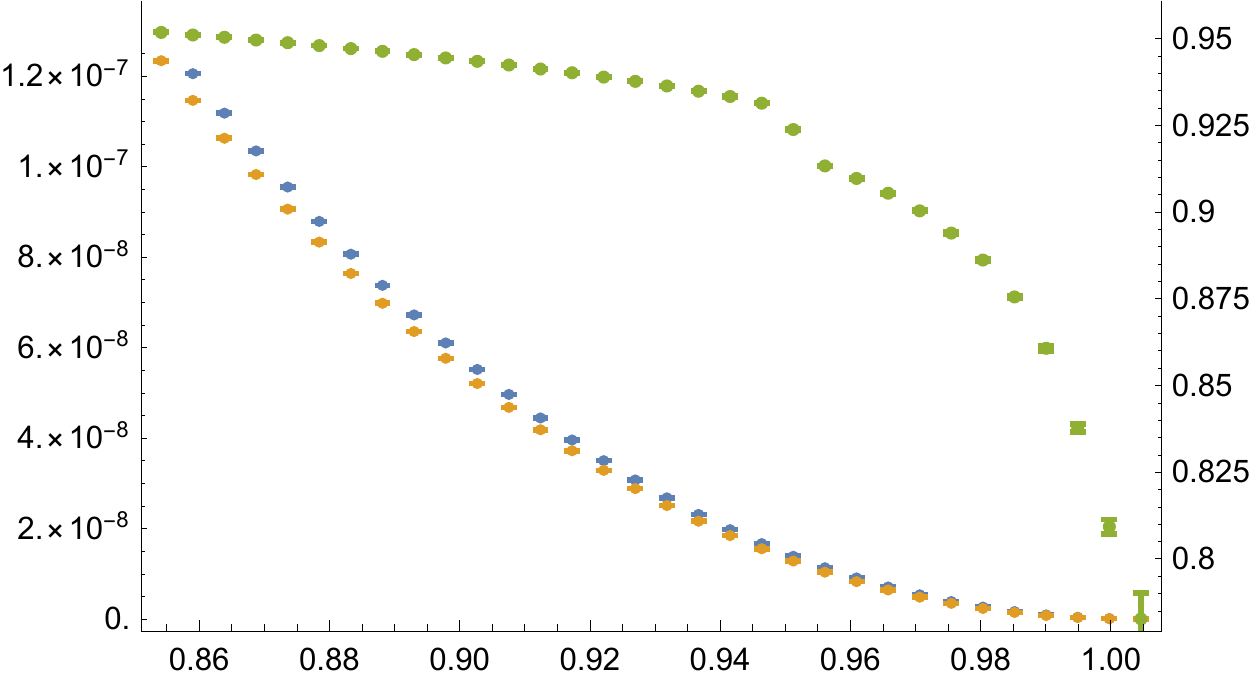}};

\node at ( 0,-0.2) {$x$};
\end{tikzpicture}
}}

  \end{minipage}
  \hfill
\begin{minipage}[b]{0.48\textwidth}

\subfigure[$\D\mathcal{B}/\D(\cos\theta_{e\gamma})$ ]{
\scalebox{0.8}{
\begin{tikzpicture}
\node [anchor=south] at (0,0) {\includegraphics[width=8.5cm]{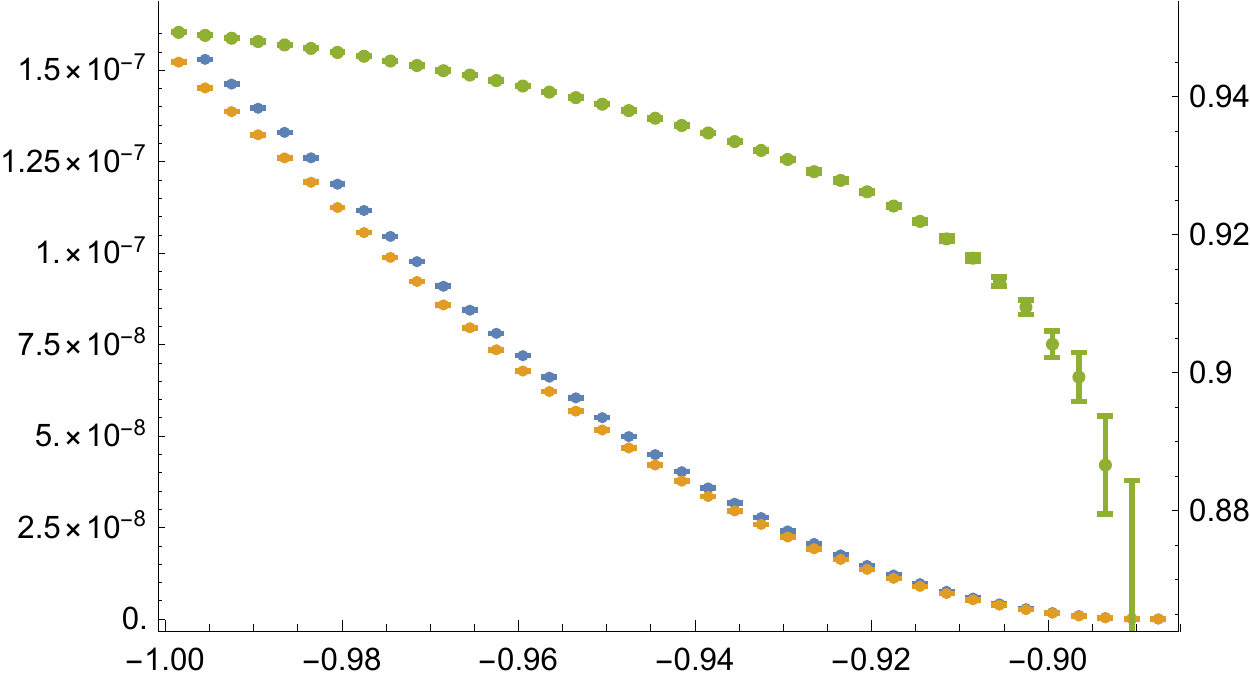}};

\node at ( 0,-0.2) {$\cos\theta_{e\gamma}$};
\end{tikzpicture}
}}

\bigskip

\subfigure[$\D\mathcal{B}/\D y$]{
\scalebox{0.8}{
\begin{tikzpicture}
\node [anchor=south] at (0,0) {\includegraphics[width=8.5cm]{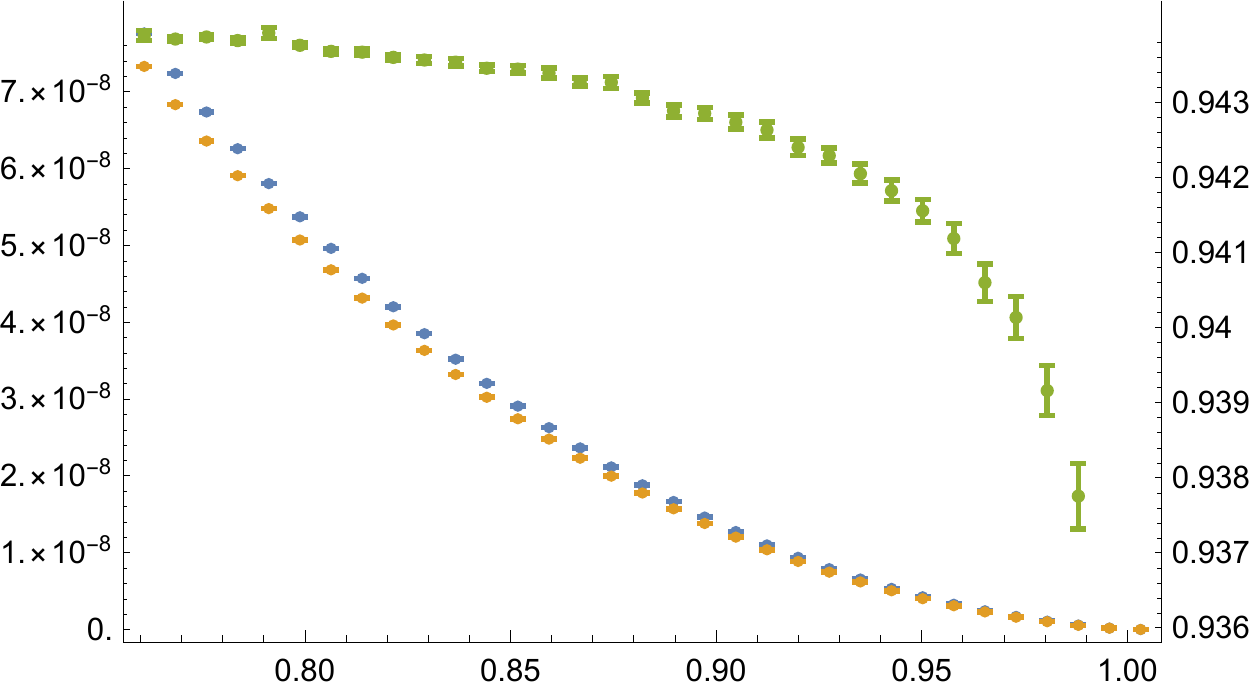}};
\node at ( 0,-0.2) {$y$};

\end{tikzpicture}
}}

  \end{minipage}

\caption{Various distributions with the MEG cuts. NLO (LO) results are
shown in orange (blue) with the scale for the observable on the
left. The differential $K$-factor is show in green with the scale on
the right. }
\label{fig:megdist}
\end{figure}

For illustrative purposes, we now consider some distributions with the
full cuts, i.e. including~\eqref{eq:cuts:geom_electron}. We start with
the missing energy distribution, $\D\mathcal{B}/\D\Em$, as it is of
special interest to the MEG experiment. In the region with very little
invisible energy  the radiative decay of the muon is an irreducible
background to the lepton-flavour violating decay $\mu\to e \gamma$
that is searched for by MEG. We define the invisible energy as
\begin{align}
\Em = m_\mu - E_e - E_\gamma\,.
\end{align}
The result is shown in the top-left panel of
Figure~\ref{fig:megdist}. For the bulk of the distribution, the
corrections are of the order of $-5\%$, but in the tail they increase
substantially. We also note that the distribution itself falls rapidly
towards zero for $\Em \to 20~\MeV$, due to the kinematic constraints.

In the top-right panel of Figure~\ref{fig:megdist} we show the
distribution of the cosine of the angle between electron and photon,
$\cos\theta_{e\gamma}$. Due to the energy cuts, the neutrino energy is
restricted to be small and the electron and the photon are nearly
back-to-back. Indeed, more than 99\% of the events lie between $-1 \le
\cos\theta_{\gamma e} \le -0.9$.  Again, the corrections are $-5\%$
for the bulk of events.

Because the energy of the electron and the photon can be measured by
the MEG experiment they, too, are of particular interest. It is
customary to show the energy fractions 
\begin{align}
x = \frac{2E_e}{m_\mu} \qquad  \mbox{and} \qquad 
y = \frac{2E_\gamma}{m_\mu}
\label{xydef}
\end{align} 
instead of the energy itself. Due to the cuts~\eqref{eq:cuts:energy}
we have $x \gtrsim 0.85$ and $y \gtrsim 0.76$.  Our NLO predictions
for the spectrum $\D\mathcal{B}/\D x$ and $\D\mathcal{B}/\D y$ are
shown in the lower panels of Figure~\ref{fig:megdist}.

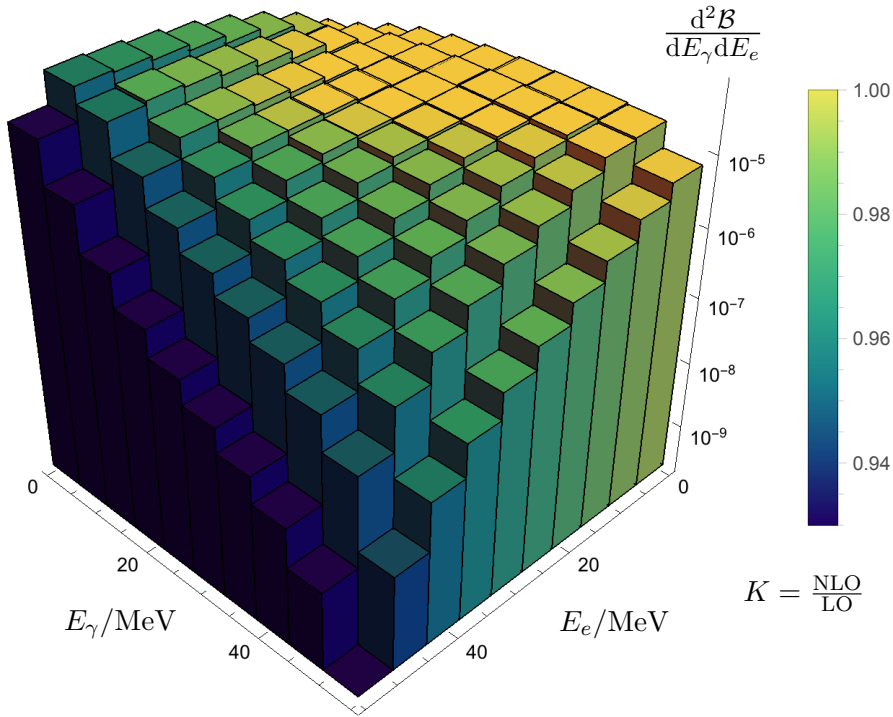
\begin{figure}[t]
\centering
\begin{tikzpicture}
\node [anchor=south] at (0,0) {\includegraphics[width=0.8\textwidth]{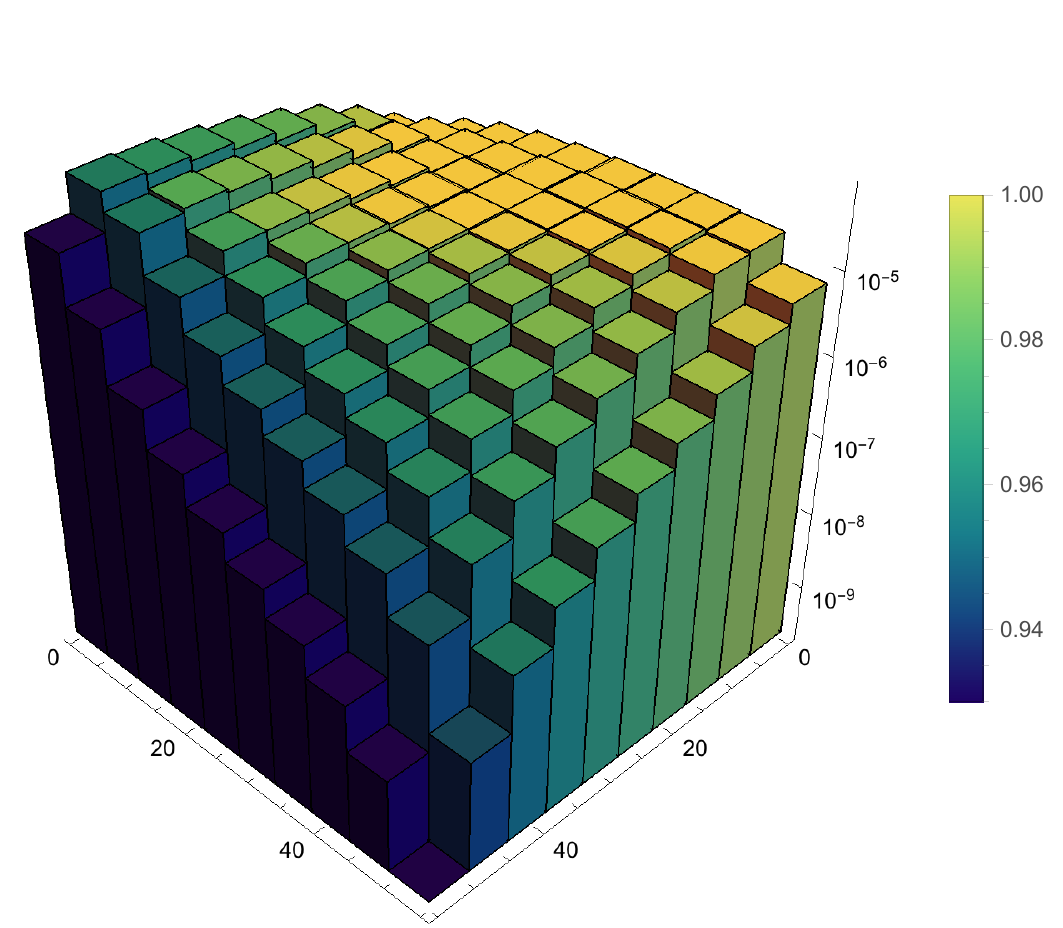}};

\node at (-4.3, 1.4) {$E_\gamma /\MeV$};
\node at ( 2.2, 1.4) {$E_e/\MeV$};
\node at ( 3.5, 9.2) {\Large$\tfrac{\D^2\mathcal{B}}{\D E_\gamma \D E_e}$};
\node at ( 4.7, 1.8) {$K=\tfrac{\text{NLO}}{\text{LO}}$};
\end{tikzpicture}
\caption{ The double differential distribution $\D^2\mathcal{B}/\D
  E_\gamma \D E_e$. The $K$ factor is shown through the colouring. A
  hard cut of $E_\gamma\ge 2\,\MeV$ was imposed on the visible photon.
}
\label{fig:meg:double}

\end{figure}

Additionally to single differential distribution we have also computed
the double differential distribution $\D^2\mathcal{B}/\D E_\gamma \D
E_e \sim \D^2\mathcal{B}/\D x \D y$ to demonstrate the impact of the
energy cuts. For this plot only the geometric MEG
cuts~\eqref{eq:cuts:geom_photon} and~\eqref{eq:cuts:geom_electron}
were used. Note that even though $E_\gamma$ starts at 0 in
Figure~\ref{fig:meg:double}, this is unphysical as a cut on the photon
energy is required to ensure infrared safety. We take $E_\gamma \ge
2\,\MeV$. To be precise, apart from an electron
with~\eqref{eq:cuts:geom_electron} we require exactly one photon with
$E_\gamma \ge 2\,\MeV$ and \eqref{eq:cuts:geom_photon}. Additional
photons must satisfy~\eqref{eq:cuts:soft}.  The distribution has the
expected sharp fall for increasing $x \sim E_e$ and $y\sim
E_\gamma$. The corrections are particularly important in the region of
large $x$, i.e. for large electron energies. This is also expected
from the bottom-left panel of Figure~\ref{fig:megdist}.

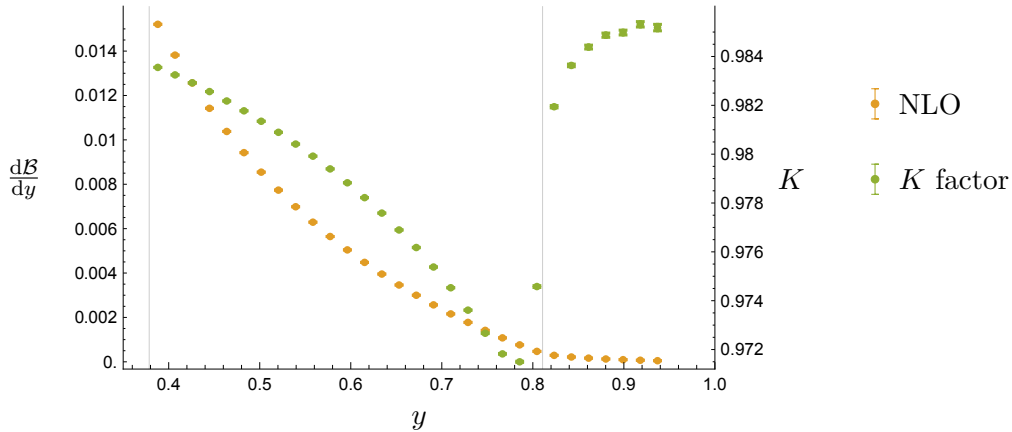
\begin{figure}[t]
\centering

\begin{tikzpicture}
\node [anchor=south] at (0,0){\includegraphics[width=0.6\textwidth]{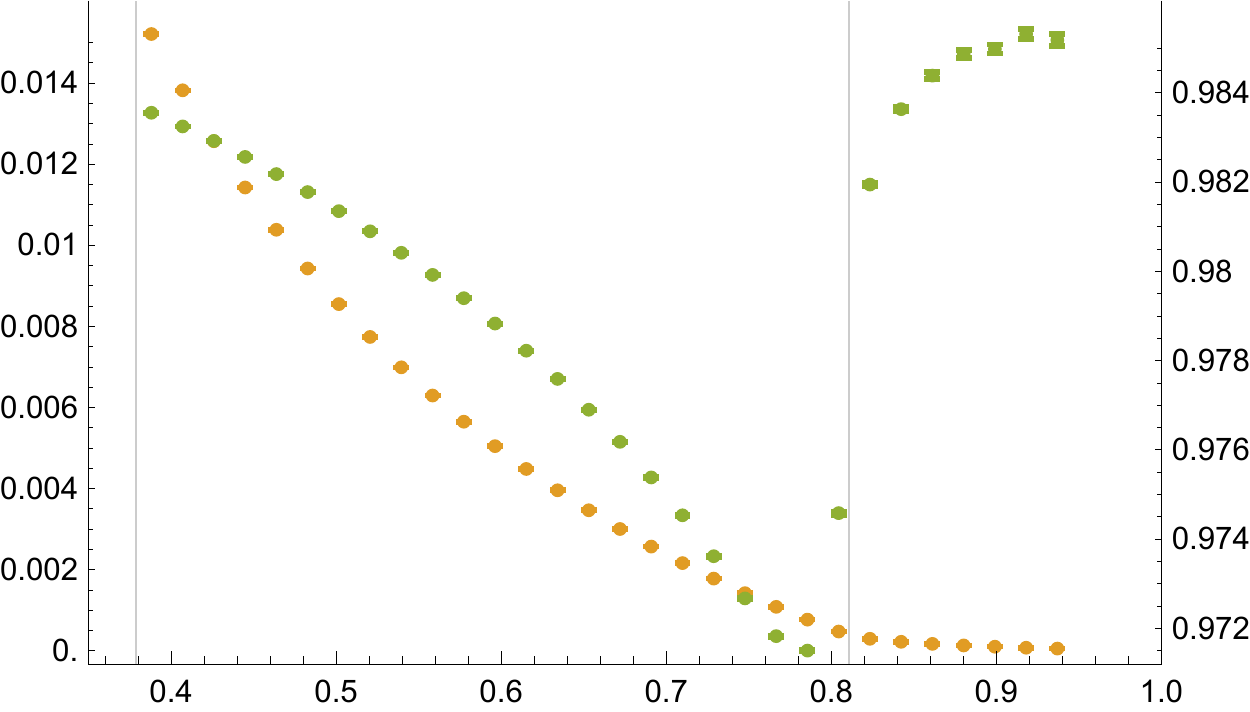}};

\node at ( 0,-0.2) {$y$};
\node at (-5.2,   3) {$\frac{\D\mathcal{B}}{\D y}$};
\node at ( 4.9,3   ) {$K$};

\legerrorpoint{math2}{6,4.0}{NLO}
\legerrorpoint{math3}{6,3.0}{$K$ factor}
\end{tikzpicture}

\caption{ The NLO differential decay distributions w.r.t. to the
  energy fraction of the photon, $y=2E_\gamma/m_\mu$, modeled for the
  Mu3e detector.  }

\label{fig:mu3e:photon}
\end{figure}

As a final illustration of the flexibility of our code, we now turn to
the Mu3e experiment~\cite{Blondel:2013ia,mu3eprivate}. In the normal running 
mode Mu3e cannot detect photons. However, it is
in principle possible to detect them should they convert into electron
positron pairs of sufficient energy. In order to mimic this situation
we consider the following cuts:
\begin{subequations}
\label{eq:mu3e}
\begin{align}
&E_\gamma > 2\,E_e\big|_{\text{min}} = 20\,\MeV\,, \qquad\qquad
E_e     > 10\,\MeV
               \label{eq:mu3e:energy}
\end{align}\begin{align}
&|\cos\theta_\gamma| < 0.8\,, \qquad\qquad
|\cos\theta_e     | < 0.8\
               \label{eq:mu3e:geom}
\end{align}\begin{align}
&E_{\gamma_2} <  \begin{cases}
20\,\MeV & \quad \text{if $|\cos\theta_\gamma|<0.8$}\\
\infty & \quad \text{otherwise}
\end{cases}
\end{align}
\end{subequations}
The cuts on $\cos\theta$ are a simplistic way to include the geometry
of the detector. We require one and only one photon in the detector,
meaning that an event with real radiation of more than the threshold
of $20\,\MeV$ will be discarded if $|\cos\theta_\gamma|<0.8$. As for
the MEG results, we set the polarization $\vec P_\mu = -0.85\ \vec z$.
As a demonstration we show the photon energy distribution
$\D\mathcal{B}/\D y$ in Figure~\ref{fig:mu3e:photon}.  We refrain from
showing the LO curve, as it is basically indistinguishable from the
NLO curve. In accordance with the fact that the cuts \eqref{eq:mu3e}
are not very restrictive, the size of the corrections is rather
modest. However, the $K$-factor, shown in green, is far from
universal. The kink in the $K$ factor at $y\simeq 0.8$ is due to the
cut on $E_e$. This shows once more that cuts can have a large impact
on the shape and size of NLO corrections.

\section{Conclusion}\label{sec:conclusion}

In this article we have presented a general purpose parton-level Monte
Carlo program for the NLO corrections within the Fermi theory of the
radiative decay of leptons. This generalizes an earlier NLO
computation~\cite{Fael:2015gua} of the branching ratio for these
decays in that now arbitrary cuts can be implemented. By today's
standards this is not a very complicated process and it is somewhat
surprising that such a Monte Carlo program had not been presented
earlier. Possible reasons might be that until recently, the
experimental result were not very precise and the (pure QED)
corrections could have been expected to be very modest, i.e. at the
order of 1\%.

However, in many important circumstances, the corrections are much
larger, as pointed out also in~\cite{Fael:2015gua,Fael:2016hnz}. In
regions of phase space corresponding to relatively stringent cuts, the
corrections can easily reach 10\%.  This is the case for background
studies to $\mu\to e \gamma$ searches as well as for recent
measurements of the branching ratio of radiative tau decays.

With increasing precision of the measurements~\cite{Lees:2015gea,
  Pocanic:2014mya} a fully-differential treatment at NLO is mandatory
to obtain reliable predictions. Indeed, we have shown that it seems
very plausible that the 3.5 standard deviation discrepancy between the
\textsc{BaBar} measurement of $\mathcal{B}(\tau\to e\,
\nu\bar{\nu}\gamma)$ and the NLO SM result is related to not using a
full NLO calculation when estimating the efficiency. With the program
presented here,  there is
no longer any reason not to use a full NLO calculation and we hope 
this helps in making more precise comparisons between theory and 
experiment for radiative lepton decays in the future.

\acknowledgments{We are grateful to Gionata Luisoni for his invaluable
  help while modifying GoSam and to Matteo Fael and Massimo Passera
  for comparing results and helpful discussions. Furthermore, we thank
  Angela~Papa, Giada~Rutar, Ann-Kathrin~Perrevoort, 
  Alberto~Lusiani and Dinko~Pocanic for very useful discussions
  regarding the MEG, Mu3e, BaBar and Pibeta experiments.  GMP and YU
  are supported by the Swiss National Science Foundation (SNF) under
  contract 200021\_160156 and 200021\_163466, respectively.}

\bibliographystyle{JHEP}
\bibliography{references}{}

\end{document}